\documentclass[11pt,twoside]{article}
\usepackage{asp2010}
\usepackage{natbib}

\newcommand{\mockalph}[1]{}

\aspvolume{471} 
\aspvoltitle{Origins of the Expanding Universe: 1912-1932}
\aspcpryear{2013} 
\aspvolauthor{Michael J. Way and Deidre Hunter, eds.}

\begin{document}

\title{The Contribution of V.~M. Slipher to the Discovery of the Expanding Universe}

\author{C. O'Raifeartaigh\affil{Department of Computing, Mathematics and Physics,
Waterford Institute of Technology, Waterford, Ireland}}


\begin{abstract} A brief history of the discovery of the expanding universe is
presented, with an emphasis on the seminal contribution of V.~M. Slipher. It is
suggested that Hubble's `discovery graph' of 1929 could also be known as the
Hubble-Slipher graph. It is also argued that the discovery of the expanding
universe matches the traditional view of scientific advance as a gradual process
of discovery and acceptance, and does not concur with the Kuhnian view of
science progressing via abrupt paradigm shifts.
\end{abstract}

\section{Introduction}

The discovery of the expanding universe marks one of the great
advances of 20th century science. It lies at the heart of today's cosmology and
forms a cornerstone of the evidence underpinning the modern `big bang' model of
the origin of the universe. Several comprehensive accounts of the discovery are
available
\citep{north1965measure,smith1982expanding,kragh1999cosmology,nussbaumer2009discovering};
however, the seminal contribution of V.~M. Slipher
remains relatively unknown to the scientific community and to the wider public. 

A brief overview of the discovery of the expanding universe is
presented, with an emphasis on Slipher's contribution. The review is presented
as distinct narratives of theory and observation, as much of the key
astronomical work was carried out independently of emerging theory. From the
analysis, we conclude that `Hubble's law' is a reasonable name for an empirical
relation between velocity and distance for the spiral nebulae, but suggest that
Hubble's `discovery graph' of 1929 could alternately be known as the
Hubble-Slipher graph. We also argue that the brief history presented matches the
classic view of scientific progress as a slow, cumulative process of theory and
experiment, followed by a long period of persuasion and gradual acceptance, and
does not support a view of science progressing by an abrupt transition to a new
paradigm, as suggested by Thomas \cite{kuhn1996structure}.

\section{A brief history of observation }

 In 1909, Vesto Melvin Slipher, a young astronomer working at the
Lowell Observatory in Flagstaff, Arizona, was set the task of studying the
spectrum of light from the Andromeda nebula. The motivation for this study was
the belief among many astronomers that the spiral nebulae constituted solar
systems in early stages of evolution. In particular, Percival Lowell, the
founder and director of the observatory, hoped that a study of the spiral
nebulae might yield important information about the origins of our own solar
system \citep{hoyt1976lowell}. For this work, young Slipher had at his
disposal a 24-inch refracting telescope by Alvan Clark and a spectrograph made
by John Brashear \citep{1927PASP...39..143S}.

 The use of spectroscopy to study the composition and motion of
celestial objects was an established tool in astronomy by this time. In
particular, the measurement of the velocity of stars by means of the Doppler
effect had been well established by observers such as William Wallace Campbell
at the Lick Observatory \citep{1906PASP...18..307C}. In this
effect, the spectral lines of light emitted by an object moving towards an
observer are measured by the observer as shifted in frequency towards the higher
(or blue) end of the spectrum, and shifted towards the lower (or red) end if the
object is moving away. However, the study of the spectra of the spiral nebulae
had proved problematic even for the world's largest telescopes, due to the
faintness of their light. Experienced astronomers such as Julius Scheiner and
Max Wolf at the Heidelberg Observatory and Edward Fath at the Lick Observatory
had obtained spectrograms that suggested the spirals contained stellar systems,
but the images were not clear enough to study the spectral lines in detail
\citep{1899ApJ.....9..149S,1909PASP...21..138F, 1912SHAWA...3.....W}.
Thus, Slipher set about the task
with some trepidation \citep{hoyt1980vesto}. Experimenting carefully
over many months, he found that good spectra of the nebulae could be obtained
using a spectrograph fitted with a camera lens of very short focus, a prism of
high angular dispersion and a wide collimator slit. His key discovery was that
the results depended critically on the speed of the spectrograph, rather than
the aperture of the telescope \citep{hoyt1980vesto}. Thus, useful
measurements of the faint nebulae could be carried out at the relatively modest
telescope at the Lowell Observatory.

 In September 1912, Slipher obtained the first clear spectrum of
Andromeda, and by January 1913, he had four plates on which the spectral lines
of the nebula were clearly visible. His analysis of the plates gave a surprising
result; the spectral lines were significantly blue-shifted, suggesting that the
spiral was approaching at a radial velocity of 300 km s$^{-1}$
\citep{1913LowOB...2...56S}. This was the first measurement of the velocity of a
spiral nebula and it was greeted with some skepticism because it was much larger
than the known velocities of stars \citep{Campbell-1913}. However,
the measurement was soon confirmed by well-known astronomers such as William H.
Wright at the Lick Observatory and Francis Pease at Mt. Wilson
\citep{1915PASP...27..134P}. 

\begin{figure}[ht]
\center{\includegraphics[scale=0.4]{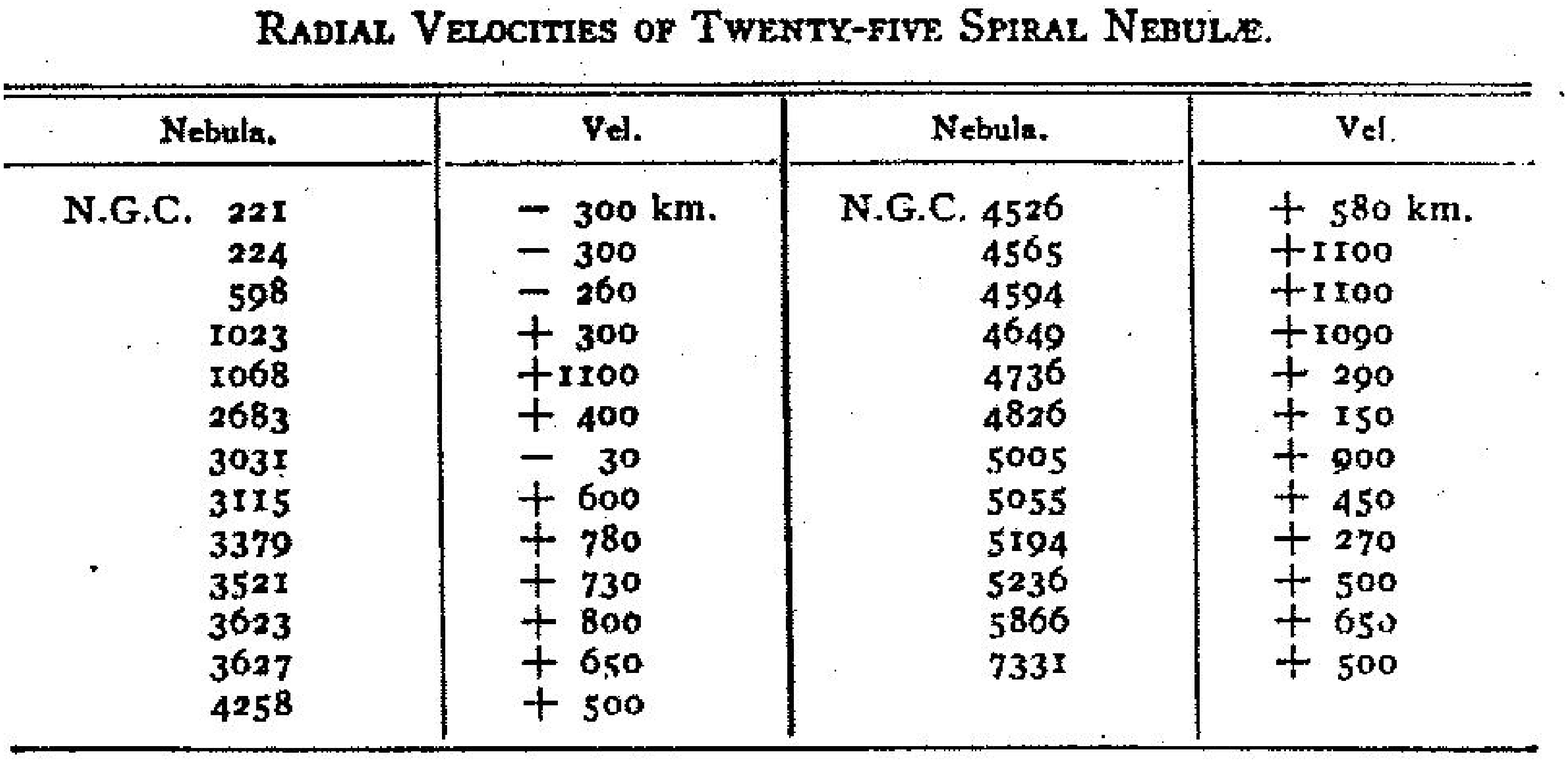}}
\caption{Radial velocities in km s$^{-1}$  of 25 spiral nebulae published by
V.~M. Slipher in 1917. Negative terms indicate velocities of approach while
positive velocities are receding.}\label{oraifeartaighfig01}
\end{figure}

By 1917, Slipher had measured spectra for 25 spiral nebulae
\citep{1917PAPhS..56..403S}. Of these, four were blue-shifted
(indicative of a radial velocity towards the observer) and the remainder were
red-shifted, indicative of objects receding from the observer. Of particular
interest were the speeds of recession, ranging from 150 to 1100 km s$^{-1}$
(see Fig. \ref{oraifeartaighfig01}).
Such large recession velocities were a great anomaly and suggested to some
that the spirals could not be gravitationally bound by the Milky Way. Thus,
Slipher's redshift observations became well-known as one argument for the
`island-universe' hypothesis, the theory that the spiral nebulae constituted
distinct galaxies far beyond the Milky Way.\footnote{In fact,
Slipher's argument was rather more subtle. He derived a
mean velocity of 700 km s$^{-1}$ for the Milky Way galaxy from his observations of the
spirals, from which he concluded that the nebulae were similar astronomical
objects.}
As he put it himself,
\emph{``It has for a long time been suggested that the spiral nebulae are
stellar systems seen at great distances. This is the so-called ``island
universe'' theory, which regards our stellar system and the Milky Way as a great
spiral nebula which we see from within. This theory, it seems to me, gains
favour in the present observations''}
\citep{1917PAPhS..56..403S}. However, the debate could not be settled until
the distances to the spirals had been measured.

Meanwhile, Slipher continued his spectrographic observations of the
nebulae. By 1922, he had amassed radial velocities for 41 spirals, almost all of
which were red-shifted. Unfortunately, he did not formally publish the full
dataset in a journal; they became known to the community when they were
published in an early textbook on general relativity
\citep{eddington1923mathematical} and in a paper by the astronomer Gustav
\cite{1925ApJ....61..353S}.

The problem of measuring the distances to the spiral nebulae was
solved by the astronomer Edwin Hubble in the 1920s.\footnote{
The astronomer Ernst \"{O}pik was the first to give a reliable estimate of
the distance to a spiral nebula \citep{1922ApJ....55..406O}. However,
he used a theoretical method that was not appreciated for many years.}
Working at the
world's largest telescope, the 100-inch Hooker reflector at the Mt. Wilson
Observatory, Hubble was able to resolve stars known as \emph{Cepheid
variables} in three of the nebulae. Such stars have the unusual property that
their intrinsic luminosity can be determined by measuring a periodic variation
in their brightness, a phenomenon that was first discovered by Henrietta Leavitt
of the Harvard College Observatory \citep{1908AnHar..60...87L}, and
developed into a powerful technique for measuring stellar distance by Ejnar
Hertzsprung and Harlow Shapley
\citep{1913AN....196..201H, 1918ApJ....48...89S}. Hubble's observations of Cepheids
in three nebulae allowed him to measure the distance to those spirals, and the
results indicated that they lay far beyond the limits of the Milky Way, settling
the `island universe' debate at last \citep{1925Obs....48..139H,1926ApJ....64..321H}. 

\begin{figure}[ht]
\center{\includegraphics[scale=0.4]{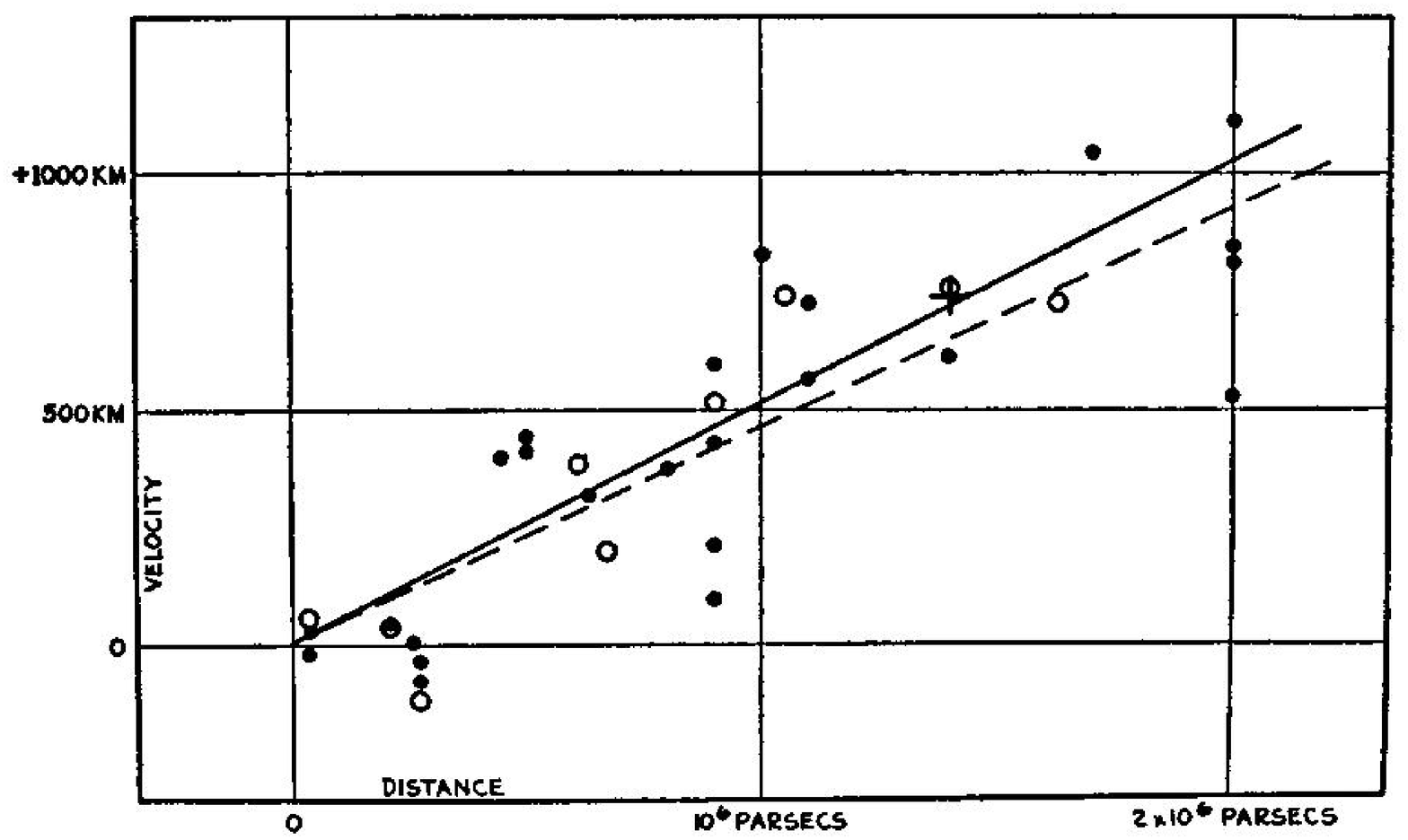}}
\caption{Graph of radial velocity versus distance for the spiral
nebulae. Reproduced from \cite{1929PNAS...15..168H}. Black data
points represent 24 individual nebulae; almost all but four of the velocities are from
Slipher, as listed by Eddington in 1923.}\label{oraifeartaighfig02}
\end{figure}

The confirmation that the spiral nebulae are distinct galaxies far
beyond our own led to renewed interest in the puzzle of Slipher's redshifts. The
next step was to investigate whether there was a simple relation between the
distance to a given galaxy and its velocity of
recession.\footnote{There were several early attempts to establish a relation between
velocity and distance for the nebulae, notably by Knut Lundmark in 1924; however
the distances to the spirals were not well established at this point.}
By 1929, Hubble had amassed reliable estimates of the distances to 24 spirals;
combining these with the corresponding redshifts from Slipher,
and four redshift measurements acquired at Mt. Wilson by his assistant Milton Humason,
Hubble obtained the velocity/distance
graph shown in Fig. \ref{oraifeartaighfig02}.
Despite considerable scatter, a linear relation between
radial velocity and distance was discernible. Hubble calculated a value of 500
km s$^{-1}$ Mpc$^{-1}$ for the slope of the solid line shown, and noted that it was consistent
with preliminary studies of a more distant nebula (a spiral of velocity 3995
km s$^{-1}$ at an estimated distance of 7 Mpc). The graph was published in the
prestigious \emph{Proceedings of the National Academy of Sciences} and it became very
well known \citep{1929PNAS...15..168H}. Unfortunately, Hubble did not
acknowledge his use of Slipher's velocity measurements in the paper, and this is
perhaps one reason the result later became known as Hubble's law.

By the time the graph of Fig. \ref{oraifeartaighfig02} was published, Hubble had embarked
on a program to extend the study to even more distant nebulae. Using a
state-of-the-art spectrograph with a specially designed `Rayton' camera lens in
conjunction with the great 100-inch reflecting telescope at Mt. Wilson, he and
Humason measured distances and redshifts for forty more spirals, demonstrating a
linear relation between velocity and distance out to a distance eighteen times
that of Fig. \ref{oraifeartaighfig02} \citep{1931ApJ....74...43H}.\footnote{In
the 1931 study, the distances of the nebulae were estimated from their
apparent magnitudes, as individual stars could not be resolved.}

It should not be concluded from this section that ``Hubble discovered
the expanding universe'', as is sometimes stated in the popular literature. Such
a statement confuses \emph{observation} with \emph{discovery}, as a linear
relation between recessional velocity and distance for the distant galaxies does
not in itself suggest an expanding universe. It is much more accurate to say
that the 1929 graph provided the first experimental evidence in support of the
hypothesis of an expanding universe. But what was this hypothesis?

\section{A brief history of theory}

The publication of Einstein's general theory of relativity in
1915 led to a new view of the force of gravity; according to relativity,
gravity was not an `action at a distance' between two objects, but a curvature
of space and time caused by matter or energy
\citep{1915SPAW.......844E,1915SPAW.......778E}. Thus the earth does not
interact directly with the sun, but follows a path in a space that has been
warped by the sun's great mass. This new view of gravity, space and time led
theorists to a number of mathematical models for the universe as a whole. 

 Einstein himself attempted the first model
\citep{1917SPAW.......142E}. Assuming a uniform distribution of matter on the
largest scales, he discovered that relativity predicts a universe that is
\emph{dynamic}, i.e. whose radius expands or contracts in time. Like most
scientists of the day, Einstein presumed that the universe is \emph{static,}
i.e. unchanging in time (no astronomical evidence to the contrary was known at
this point). Given that gravity is an attractive force that could cause the
universe to contract, he added a small term to his equations that could
counterbalance the effect, a term he named the `cosmological constant.'
This analysis led Einstein to a model of the cosmos that is static in time and
of closed spatial geometry - a finite universe whose radius could be calculated
from the density of matter.\footnote{Constants of integration occur
naturally in the solution of differential
equations such as the Einstein field equations. The size of the cosmological
constant is constrained by the requirement that relativity predicts the motion of
the planets, but there is no reason it should be exactly zero. Einstein's
suggestion was that a small, non-zero constant of integration could
simultaneously render the universe static and give it a closed curvature, neatly
removing the problem of boundary conditions. It was later shown that this
solution is in fact unstable \citep{1930MNRAS..90..668E}.}

The Dutch astronomer Willem de Sitter also applied Einstein's field
equations to the cosmos. Assuming a universe empty of matter, de Sitter found a
second solution that also appeared to be static
\citep{1917MNRAS..78....3D}. A curious feature of the model was the
prediction that any matter introduced into this universe would recede from the
observer, observable as a redshift. A second redshift effect due to an apparent
slowing down of atomic vibrations was also predicted. The `de Sitter effect'
became quite well-known in the 1920s, and astronomers such as Ludwig
Silberstein, Carl Wirtz, Knut Lundmark and Gustav Str\"{o}mberg sought to measure
the curvature of space from the redshifts of stars, global clusters, planetary
and spiral nebulae 
\citep[e.g.][]{1924MNRAS..84..363S,1924AN....222...21W,1924MNRAS..84..747L,1925ApJ....61..353S}.
In general, these attempts to match theory with
observation were not successful, due to a flaw in de Sitter's analysis (see
below). However, it's worth noting that Lundmark provided the first
velocity/distance plot for the \emph{spirals}, although the results were not very clear
\citep{1924MNRAS..84..747L}. A description of the work of
Silberstein, Wirtz, Lundmark and Str\"{o}mberg and can be found in the essay by
Harry Nussbaumer in this book.

In 1922, the Russian theoretician Alexander Friedman published
solutions to the Einstein field equations that included not only the static
solutions of Einstein and de Sitter, but also a universe of time-varying radius
\citep{1922ZPhy...10..377F}. In the language of relativity, he was
the first to allow the possibility of a \emph{dynamic} space-time metric for
the universe. With another paper in 1924, Friedman established almost all the
main possibilities for the evolution of the cosmos and its geometry
\citep{1924ZPhy...21..326F}, an analysis that provides the framework for the
models of today. However, little attention was paid to Friedman's work at the
time because most people, including Einstein, considered time-varying models
of the universe to be unrealistic.\footnote{Einstein first accused
Friedman of a making a mathematical error \citep{1922ZPhy...11..326E}.
He later withdrew the comment \citep{1923ZPhy...16..228E}, but a draft
copy of his retraction contains the revealing phrase \emph{`denen eine
physikalische Bedeutung kaum zuzuschreiben sein d\"{u}rfte'}
or `\emph{to this a physical significance can hardly be ascribed'}
\citep{Einstein-1923}.}

Friedman himself made no
attempt to connect his theory to experiment as he was unaware of Slipher's
observations, and he died four years before the publication of the
Hubble-Slipher graph. A full discussion of Friedman's contribution is given in
the essay by Ari Belenkiy in this book.

Unaware of the earlier work of Friedman, the Belgian theoretician Georges\newline
Lema\^{i}tre also discovered that the application of Einstein's field
equations to the cosmos gives time-varying solutions. In his first contribution
to the field, he spotted a significant inconsistency in de Sitter's analysis;
correcting the error showed that de Sitter's empty universe is not static
\citep{1925Lemaitre.60..188}. A theoretician with
significant training in astronomy, Lema\^{i}tre was well aware of Slipher's
redshifts and Hubble's emerging measurements of the vast distances to the spiral
nebulae \citep{1987Cent...30..114K}. His great insight was to link the
recession of the spirals with a relativistic expansion of space-time. (Note that
as an expansion of space-time \emph{metric}, the effect would be detectable
only on extra-galactic scales, not in the earlier studies of stars, globular
clusters and planetary nebulae). In a pioneering paper in 1927, Lema\^{i}tre derived
a universe of expanding radius from Einstein's equations, and then estimated the
rate of expansion using average values of velocity and distance for the spirals
from Slipher and Hubble respectively \citep{1927ASSB...47...49L}.
He obtained a value of 575 km s$^{-1}$ Mpc$^{-1}$ for the coefficient of expansion
(as well as an alternate estimate of 625 km s$^{-1}$ Mpc$^{-1}$ using a statistical weighting
method): these values were in good agreement with Hubble's estimate two years
later (see above). Thus Lema\^{i}tre was undoubtably the first to connect the theory
of the expanding universe with observation. However, his work went unnoticed
because it was published in French in a little-known Belgian journal. Lema\^{i}tre
did little to promote his model, perhaps due to a negative reaction from
Einstein; the latter declared the expanding model `abominable', and added that
such models had in any case already been suggested by Alexander Friedman
\citep{Lemaitre1958rencontres}!

It should be noted that during these years, other theoreticians such
as Hermann Weyl \citep{weyl1993raum,weyl1919uber,weyl1923allgemeinen}, Cornelius
\cite{lanczos1922bemerkung,lanczos1923rotverschiebung}, Howard Percy
\cite{1929PNAS...15..822R} and Richard \cite{1929ApJ....69..245T,1929PNAS...15..297T}
also applied Einstein's field
equations to the study of the cosmos. All of them spotted the inconsistency in
the model of de Sitter; however, Friedman and Lema\^{i}tre were the first to make
the key step of specifically allowing time-varying solutions for the radius of
the universe. 

\section{A convergence of theory and observation}

The publication of Hubble's velocity/distance graph of 1929 did not
cause a major stir in the scientific community at large, but the relativists
paid close attention. At a seminar at the Royal Astronomical Society in January
1930, de Sitter admitted that a linear relation between distance and radial
velocity for the nebulae could not be explained in the context of his own model
or that of Einstein. In the ensuing discussion, the eminent British astronomer
Arthur Stanley Eddington suggested that a new model of the cosmos was needed.
Their discussion was published in the proceedings of the meeting
\citep{1930Obs..53....37D} and came to the attention of Lema\^{i}tre, who wrote to
Eddington to remind him of his 1927 paper. Eddington immediately grasped the
significance of Lema\^{i}tre's work and quickly made others aware of it
\citep{1930MNRAS..90..668E}. He also arranged for it
to be translated and republished in the widely-read \emph{Monthly Notices of the Royal
Astronomical Society}. The paper duly appeared
\citep{1931MNRAS..91..483L} although the section where a coefficient of expansion
is estimated from observational data was not included. It has recently been
confirmed that this revision was carried out by Lema\^{i}tre himself in the light of
Hubble's 1929 paper \citep{2011Natur.479..171L,Lemaitre-1931}.

Supported by the empirical data of Hubble's 1929 paper, Lema\^{i}tre's
model of a universe of expanding radius became widely known
\citep[e.g.][]{1931Natur.128..706D}. Einstein publicly accepted
the expanding model during a visit to the United States in early 1931, drawing
worldwide attention to the work of Hubble, Humason, Lema\^{i}tre and Tolman
\citep{1931NYT.......844E}; he also published a short academic
paper on the expanding universe later that year
\citep{1931SPAW.......235E}. Thus by the early 1930s, it seemed to many relativists
and some astronomers that an astonishing new phenomenon, the expanding universe,
had been discovered that could be explained in a natural way in the context of
the general theory of relativity. 

\section{On the naming of laws and equations}

In time, the velocity/distance graph of Fig. \ref{oraifeartaighfig02} became known as
`Hubble's law.' It is not entirely clear when or why this nomenclature became
the norm. One factor may have been Hubble's failure to acknowledge Slipher's
data in the `discovery' paper of 1929 \citep{1929PNAS...15..168H}.
Lema\^{i}tre also neglected to cite Slipher directly in his seminal 1927 and 1931
papers \citep{1927ASSB...47...49L,1931MNRAS..91..483L};
these omissions may have set a precedent for authors of subsequent papers. A
second factor may have been Hubble's well-known vigilance in defending and
promoting the contribution of Mt. Wilson astronomers,\footnote{Hubble's accusation
of plagiarism on the part of Lundmark \citep{1926ApJ....64..321H} and his
aggressive letter to de Sitter \citep{Hubble-1931} are good examples of this
attitude.} an attitude that
was in marked contrast with Slipher's reticence in such matters
\citep{hoyt1980vesto}. Indeed, it is remarkable that Slipher never formally
published the full set of his painstaking redshift measurements, but allowed
them to be circulated by Eddington and Str\"{o}mberg instead. However, the most
important factor in the naming of Hubble's law is undoubtedly one of social
context; Hubble was a famous astronomer working at the world's foremost
observatory, while Slipher was a lesser-known figure working at a smaller
facility best known for controversial claims concerning the observation of
canals on Mars.\footnote{Lowell's persistent claims of the observation of
canals on Mars damaged the reputation of the Lowell observatory \citep{hoyt1976lowell}.}
Thus, the graph of 1929 became known as `Hubble's law'
and its slope as the `Hubble constant.'

It could be argued that Hubble fully merits this recognition, given
his groundbreaking measurements of the distances to the nebulae
\citep{1925Obs....48..139H,1926ApJ....64..321H}, his combination of
distance measurements with Slipher's data to obtain the first evidence for a
velocity/distance relation \citep{1929PNAS...15..168H}, and his
subsequent extension of the relation to much larger distances
\citep{1931ApJ....74...43H}. We find this a reasonable argument;
however, in order to recognize that almost all the velocity data in the
`discovery' graph of 1929 are from Slipher, we suggest that this particular
graph could also be known as the `Hubble-Slipher graph.' As Hubble once wrote in
a letter to Slipher \emph{``I have obtained a velocity/distance relation for
the nebulae using your velocities and my distances''} 
\citep{Hubble-1953}. It is one of the great ironies of science that Hubble's
measurements of distance were later substantially revised due to significant
systematic errors,\footnote{Due to an error in the classification of Cepheid
variables, Hubble's cosmological distance ladder was later substantially revised
by Walter \cite{1956PASP...68....5B} and Allan
\cite{1958ApJ...127..513S}. Hubble's distances of 1929 may also have contained some
observational errors, as suggested in the essay by John Peacock in this
book.} while Slipher's redshift data have stood the test of
time remarkably well.

It is sometimes argued that Hubble's law should be known as the
`Hubble-Lema\^{i}tre law' \citep{farrell2006day}, or even `Lema\^{i}tre's
law' \citep{2011arXiv1106.3928B}, given the pioneering contribution of
Georges Lema\^{i}tre in 1927. We do not find this a reasonable argument simply
because Hubble's law is understood as an empirical relation between velocity and
distance for the nebulae. Lema\^{i}tre did not provide any measurements of velocity
or distance, nor did he establish the linearity of the velocity/distance
relation. Instead, he \emph{predicted} a linear relation between velocity and
distance from theory, and, assuming that such a relation existed, used average
values of observational data for the spiral nebulae to estimate a coefficient
of expansion for the universe. That this calculation was something of a
provisional `guesstimate' can be seen from the fact that it is included only as
a footnote in the 1927 paper \citep{1927ASSB...47...49L}, and not
at all in the translated version \citep{1931MNRAS..91..483L}.
Lema\^{i}tre's attitude can be clearly seen in a recently-discovered letter that
accompanied his 1931 manuscript when he states; ``\emph{I do not think it is
advisable to reprint the provisional discussion of radial velocities which is
clearly of no actual interest'' } \citep{Lemaitre-1931}.
Thus, it seems to us that to credit him with the discovery
of a velocity/distance relation for the nebulae confuses theory with
observation. As he remarked many years later in a discussion of his 1927
paper,\emph{``Naturellement, avant la d\'{e}couverte et l'\'{e}tude des amas de
nebuleuses, il ne pouvait \^{e}tre question d'\'{e}tablir la loi de Hubble'' } or
``\emph{Naturally, before the discovery and study of the clusters of nebulae,
it was not possible to establish Hubble's law}''
\citep{1952MNSSA..11..110L}.\footnote{This passage is mistranslated in a recent
paper by David \cite{2011arXiv1106.3928B}.}

The above is not to understate Lema\^{i}tre's seminal contribution; he is
recognized as the first to connect the recession of the spiral nebulae with a
relativistic expansion of space-time, an expansion that he derived himself from
the Einstein field equations. He is also recognized for his retention of the
cosmological constant; where Einstein and de Sitter quickly disposed of the term
in constructing a new model of the cosmos
\citep{1932PNAS...18..213E}, Lema\^{i}tre retained it as an important component of
cosmological models (not least because of its potential to circumvent a conflict
between the age of the universe estimated from the expansion and from the known
age of stars). This approach led Lema\^{i}tre to a model of a universe whose rate of
expansion first de-accelerates and then accelerates
\citep{1934PNAS...20...12L}, remarkably similar to the best-fit models of today.
Finally, Lema\^{i}tre's characteristic blending of theory and experiment also led
him to become the first physicist to postulate a physical model for the origin
of the universe - the `primeval atom' \citep{1931Natur.127..706L}.
It is for this model, the forerunner of today's big bang model, that he is best
known.

As regards Friedman, his time-varying solutions provided a template
for all subsequent models of the evolution and geometry of the universe. Thus it
could be said that Friedman derived the possibility of an \emph{evolving}
universe from Einstein's equations, while Lema\^{i}tre, guided by observational
data, derived an \emph{expanding} universe. (Note that our universe could one
day contract, depending on the nature and time-evolution of dark energy). Today,
the `Friedman equations' appear in the first chapter of every cosmology
textbook, as do `Friedman universes.' Indeed, the contributions of both Friedman
and Lema\^{i}tre are recognized in the naming of the
Friedman-Lema\^{i}tre-Robertson-Walker (FLRW) metric,
a fundamental tool of today's theoretical cosmology.

In summary, we note that references to `Hubble's law' and `the Hubble
constant' are to be found throughout the scientific literature, while the work
of Friedman and Lema\^{i}tre is recognized in every modern textbook on cosmology. By
contrast, Slipher's contribution seems destined to be consigned to the footnotes
of history, although his pioneering observations provided a crucial part of the
first evidence for the expanding universe. This contribution was neatly
summarized by the President of the Royal Astronomical Society in 1933, in his
closing remarks on the occasion of the awarding of the society's Gold Medal to
Slipher \citep{1933MNRAS..93..476S}:

\begin{quote}
In a series of studies of the radial velocities of these
island galaxies, he laid the foundations of the great structure of the expanding
Universe, to which others, both observers and theorists, have since contributed
their share. If cosmogonists today have to deal with a universe that is
expanding in fact as well as in fancy, at a rate which offers them special
difficulties, a great part of the initial blame must be borne by our
medallist.
\end{quote}

\section{A note on paradigm shifts in science}

By the early 1930s, a new phenomenon, the expansion of the universe,
had been observed that could be explained in the context of the general theory
of relativity. In \emph{retrospect,} this fusion of theory and experiment
marked a watershed in modern cosmology, and it was a key step in the development
of today's `big bang' model of the origin for the universe.

However, the scientific community did not shift to a new view of the
universe overnight. In fact, it was many years before most physicists accepted
that the redshifts of the spiral nebulae truly represented recessional
velocities, and that an explanation for the recession could be found in terms of
a relativistic, expanding universe 
\citep{north1965measure,2003HisSc..41..141K}. During this time, many
alternate models were considered.

One such model was the `tired light' hypothesis of Fritz Zwicky. In
this theory, the redshifts of the nebulae were not due to an expansion of space,
but to a loss of energy as starlight travelled the immense distance to earth
\citep{1929PNAS...15..773Z}. Many scientists took the theory
seriously, although it was later ruled out by experiment. Other non-relativistic
models of the universe emerged 
\citep{1930MNRAS..91..128M, 1933ZA......6....1M}, and astronomers carried out many
observations in order to test the models (North 1965, chapter 11; Kragh 1999,
chapter 7). Thus, it could not be said
that, after 1931, astronomical results were interpreted in terms of one model
only, the relativistic expanding universe. 

For example, it's worth noting that Hubble declined to interpret the
velocity/distance relation in terms of an expanding universe throughout his
life. This is not to say that he was unaware of relativistic models of the
cosmos; in his paper of 1929, he remarks that \emph{``the outstanding
possibility is that the velocity/distance relation may represent the de Sitter
effect}, \emph{and hence that numerical data may be introduced into
discussions of the general curvature of space}''
\citep{1929PNAS...15..168H}. However, in subsequent years, Hubble declined to
interpret his empirical data in the context of any particular model, in case the
theory might later prove wanting \citep{hubble1958realm}. This
approach was quite common among professional astronomers at the time,
particularly in the United States 
\citep{2003HisSc..41..141K}, and it is somewhat in conflict with the modern
hypothesis of the `theory-ladeness' of scientific observation
\citep{hanson1958patterns}.

We also note that our narrative does not match Thomas Kuhn's view of science
progressing via long periods of `normal science' interspersed by relatively abrupt\newline
`paradigm shifts' \citep{kuhn1996structure}.
Instead of an abrupt transition to a new cosmological paradigm
incommensurate with the old,\footnote{The concept of `incommensurability' refers
to Kuhn's belief that a new scientific paradigm cannot be meaningfully
compared with previous models, because the underlying assumptions of
the worldviews are different \cite{kuhn1996structure}.} there was a long period
from 1930-1960 when many models were considered, as described above. Secondly, a
paradigm shift to a relativistic, expanding universe might have been expected to
trigger a great upsurge of interest in cosmology, relativity and the expanding
universe. Nothing of the kind happened; it can be seen from the citation record
\citep{marx2010accurately} that few in the wider physics
community took an interest in the expanding universe in the years 1930-1960,
despite exciting developments such as Lema\^{i}tre's `primeval atom'
\citep{1931Natur.127..706L} or the fiery infant universe of Gamow, Alpher and
Herman \citep{1948Natur.162..680G,1948Natur.162..774A}.
It seems likely that this indifference is linked to an
overall decline of interest in general relativity. Physicists found the new
theory very difficult mathematically and, outside of cosmology, it made few
predictions that differed significantly from Newtonian physics. In consequence,
the study of general relativity became consigned to mathematics departments,
with little interest from physicists and astronomers (Eisenstaedt 1989;
Eisenstaedt 2006, chapter 15).
Where a paradigm shift to the notion of a relativistic, expanding universe might
have been expected to cause a great upsurge in the study of general relativity,
the opposite happened; cosmology remained a minority sport within the physics
community for decades (North 1965, chapter 11; Kragh 1999, chapter 7; Kragh 2006,
chapter 3), a situation that did not change until the discovery of the cosmic microwave
background in 1965.\footnote{The discovery of a ubiquitous cosmic background
radiation of extremely
long wavelength offered strong support for the hypothesis of a universe that has
been expanding and cooling for billions of years.}

In conclusion, one can describe the discovery of the expanding
universe as a slow, parallel emergence of theory and observation, with many
false starts, wrong turns and re-discoveries. Once accepted among a small band
of relativists, the discovery experienced an equally slow acceptance among the
wider physics community, with alternate models being considered for many years.
This behavior mirrors the traditional model of scientific discovery as a
quasi-linear process of gradual evolution and persuasion, and does not match the
Kuhnian view of an abrupt shift to a new paradigm that becomes
incommensurate with the old. This point shall be discussed further in a
forthcoming paper.

\acknowledgements

The author would like to thank Dr. Michael Way of the NASA Goddard
Institute for Space Studies, Dr. Simon Mitton of St. Edmund's College of
Cambridge University and Dr. Joseph Tenn of Sonoma State University for many
helpful suggestions. He would also like to thank the Dublin Institute of Advanced
Studies for access to the Collected Papers of Albert Einstein (Princeton
University Press).

\nocite{1989ehgr.conf..277E}
\nocite{eisenstaedt2006curious}
\nocite{kragh2006conceptions}

\bibliography{oraifeartaigh}

\end{document}